\documentclass[aps,12pt,tightenlines,amsmath,amssymb,showpacs]{revtex4}
\usepackage{graphicx}
\usepackage{dcolumn}
\usepackage{bm}
\def\Eqref#1{Eq.~(\ref{#1})}
\newcommand{\ba}{\begin{array}}\newcommand{\ea}{\end{array}}
\font\capsten=cmcsc10 scaled\magstep0
\newcommand{\be}{\begin{equation}}\newcommand{\ee}{\end{equation}}

\def\Acal{{\cal A}}
\def\obsrep{observable representation}
\def\Rs{S}
\def\diag{\mathop{\rm diag}\nolimits} 

\def\erset{\mathbb R}
\def\Abf{{\bf A}}

\begin{document}
\title{\hfill$\hbox{\rm\small To appear in J. Phys. A}$\\~\\
Imaging geometry through dynamics: the observable representation}
\author{Bernard Gaveau}
\affiliation{Laboratoire analyse et physique math\'ematique, 14 avenue F\'elix Faure, 75015 Paris, France}
\email{gaveau@ccr.jussieu.fr}

\author{Lawrence S. Schulman}
\affiliation{Physics Department, Clarkson University, Potsdam, New York
13699-5820, USA}
\affiliation{Service de Physique Th\'eorique, CEA Saclay, 91191 Gif-sur-Yvette Cedex, France}
\email{schulman@clarkson.edu}

\author{Leonard J. Schulman}
\affiliation{Caltech, 1200 E. California Blvd. MC 256-80, Pasadena CA 91125, USA}
\email{schulman@caltech.edu}

\date{\today}
\begin{abstract}
For many stochastic processes there is an underlying coordinate space, $V$, with the process moving from point to point in $V$ or on variables (such as spin configurations) defined with respect to $V$. There is a matrix of transition probabilities (whether between points in $V$ or between variables defined on $V$) and we focus on its ``slow'' eigenvectors, those with eigenvalues closest to that of the stationary eigenvector. These eigenvectors are the ``observables,'' and they can be used to recover geometrical features of~$V$. 
\end{abstract}
\maketitle

\section{\label{sec:intro} Introduction}

For many stochastic processes there is an underlying coordinate space, $V$. For Brownian motion on $V$ the stochastic dynamics senses the structure of $V$ through the transition rules. For stochastic dynamics of (say) a spin variable, a Markov process on $2^V$, if the energy function associated with the dynamics is local and if the spin-flip rules are local there are again ways in which the coordinate space can be felt.

In this article we demonstrate how in the \textit{observable representation}, introduced in \cite{firstorder, grains, dynamicalmetric, multiplephases}, the space $V$ can be made manifest. For example, for Brownian motion on a (discretized) circle, the \obsrep, which is a plot based on eigenvectors of the matrix of transition probabilities, gives, simply, a circle. Our work may be viewed as part of a general effort to use spectral properties of operators to recover geometrical features of an underlying space. The book of Morrow and Kodaira \cite{kodaira} shows how the slowest eigenfunctions of the Laplacian on an algebraic manifold can be used to parametrize that manifold. Recent work of Coifman et al.\ \cite{coifman} has provided specific examples of the use of harmonic analysis to illustrate geometric structure. It is interesting how researchers with different goals have converged on a common technique. Our original objective was the understanding of nonequilibrium statistical mechanics, of phase transitions and the clarification of certain fundamental questions about the second law of thermodynamics. Other researchers give the geometric organization of graphs as a primary goal. Nevertheless, in both cases spectral analysis of a stochastic process has become the method of choice.

Although much of our previous work focused on the metric structure in the \obsrep, the present article is concerned more with topological features (so we have a continuum limit in mind) and is closer to the works just cited. The exploratory aspect of much of what we here present, however, allows us to study issues and spaces that as far as we know have not been previously considered. For example, we mix manifolds of different dimension (line attached to a plane), we consider power sets of a set and perhaps of most relevance to the general representation of graphs, consider cases where degeneracy or the high multiplicity of eigenfunctions leads to the need for a high dimensional representation in order to faithfully represent an underlying space. Nevertheless, we do use metric structure (as in our recalling of the barycentric coordinate theorem \cite{multiplephases} in Sec.\ \ref{sec:kawasaki}), and we also establish a new inequality relating dynamics to distance in the \obsrep\ (\Eqref{eq:bound}).

Sec.\ \ref{sec:notation} introduces the framework and notation. Following that, in Sec.\ \ref{sec:illustrationsone} we present many examples  of spaces, $V$, for which the method provides faithful representations. These include lines connected in various ways, for example non-planar graphs for which the \obsrep\ automatically requires an additional dimension for fidelity. There are also spaces of higher dimensions, a decorated plane and a torus. The next Section, \ref{sec:illustrationstwo}, deals with stochastic process where there is an internal variable for each point of $V$ (specifically an Ising model), and once again, under the right circumstances the geometry of $V$ can be discerned. For some systems the \obsrep\ demands yet higher dimension and these are studied in Sec.\ \ref{sec:proliferate}. Following that, Sec.\ \ref{sec:rationale} gives a dynamical bound on distances in the \obsrep. Finally, in Sec.\ \ref{sec:discussion} is a brief discussion of the results.

\section{\label{sec:notation}The observable representation and associated notation}

States of the system are given by $x,y\in X$, with $X$ a space of cardinality $N<\infty$. The system moves in discrete time according to a stochastic matrix $R$ defined as follows
\be
R_{xy}=``\Pr\bigl(x\leftarrow y\bigr)"=\Pr\left[\hbox{State at time $(t+1)$ is}~x\mid \hbox{State at time $t$ is}~y\right]\;.
\ee
Although many of our results are more general we here restrict attention to irreducible, detailed balance processes.

With detailed balance, the eigenvalues of $R$ are real and can written $1=\lambda_0 >|\lambda_1| \geq|\lambda_2|\dots\geq0$. There is a strictly positive stationary distribution, $p_0$, which is a right-eigenvector of $R$ with eigenvalue 1; correspondingly there is a constant left eigenvector, $A_0\equiv1$ whose eigenvalue relation \hbox{($A_0=A_0R$)} expresses conservation of probability. Because of detailed balance, $R$ can also be written using a symmetric matrix that we denote $\Rs$. Define $\sigma\equiv \diag\left(\sqrt{p_0}\right)$ \cite{note:diag}. Then $\Rs=\sigma^{-1}R\sigma$. Call the eigenvectors of $\Rs$, $\psi_\alpha$; then for a particular eigenvalue $\lambda_\alpha$ the left (``$A_\alpha$'') and right (``$p_\alpha$'') eigenvectors are related to $\psi_\alpha$ as follows
\be
p_\alpha(x)=\sqrt{p_0(x)}\psi_\alpha(x)\quad \hbox{and}
\quad A_\alpha(x)=\psi_\alpha(x)/\sqrt{p_0(x)}
\label{eq:eigvecrelation}
\ee
We take the $\psi_\alpha$'s to be orthonormal, inducing a different normalization for the $A_\alpha$'s from that used in \cite{multiplephases, firstorder, grains, dynamicalmetric}. This normalization is used in Sec.\ \ref{sec:rationale}, although the figures use the earlier normalization. For the examples given there is no substantial difference.

The $m$-dimensional \textit{observable representation} is the following collection of points \be
\Acal\equiv \left\{ {\Abf}\in\erset^m 
\mid \Abf=(A_1(x),A_2(x),\dots,A_m(x)) \hbox{~for~} x\in X\right\}
\label{eq:observrep} \,.
\ee
To visualize this, take the left eigenvectors to be row vectors. Write the first $m$ eigenvectors one atop the other. The points of the observable representation are then obtained as the \textit{columns} in the following array:
\def\lift#1{\noalign{\vskip-#1pt}}
\be
\begin{array}{cc}
~~~&\hskip 1.in\hbox{{\small Points of the observable representation} ($\downarrow$)}\cr
&\begin{array}{ccccccccc}
   &&         &\Abf(x_1)  &\Abf(x_2)  &\Abf(x_3)  &\dots  &\Abf(x_N) &~\hskip.5in~ \cr
   &&         &\downarrow &\downarrow &\downarrow &       &\downarrow \cr        
   ~~~~~~~~~~~~~~~~~~\cr
\hbox{{\small Eigenvectors}}
   & A_1\to&~~  &A_1(x_1)   &A_1(x_2)   &A_1(x_3)   &\dots  &A_1(x_N)   \cr
\hbox{($\to$)}   
   & A_2\to  &&A_2(x_1)   &A_2(x_2)   &A_2(x_3)   &\dots  &A_2(x_N)   \cr
   & \ba{ccc}\lift7\dots\cr\lift{10}\dots\cr\lift8\dots\ea
             &&\vdots     &\vdots     &\vdots     &\vdots &\vdots    \cr
   & A_m\to  &&A_m(x_1)   &A_m(x_2)   &A_m(x_3)   &\dots  &A_m(x_N)
\end{array}
\end{array}
\label{eq:visualization}
\ee
The set $\Acal$ is produced by putting the $N$ points, $\Abf(x_k)$, in $\erset^m$. In \cite{multiplephases} we focused on the properties of $\Acal$ when there was a phase transition in the system, which, as shown in \cite{firstorder}, corresponds to eigenvalue near-degeneracy for the slowest eigenvectors (those with eigenvalue magnitude nearest to 1). Thus if $\lambda_m$ is very close to 1, while $\lambda_{m+1}$ is not, $\Acal$ is a simplex with states in the phases at the vertices. States that asymptotically tend to the phases lie in the interior. The barycentric coordinates of these points with respect to the nodes of the simplex provide the probabilities that these points end in one or another phase.

\subsection*{\label{sec:prelude}Prelude}

\begin{figure}
\centerline{\includegraphics[height=.35\textheight,width=.5\textwidth]{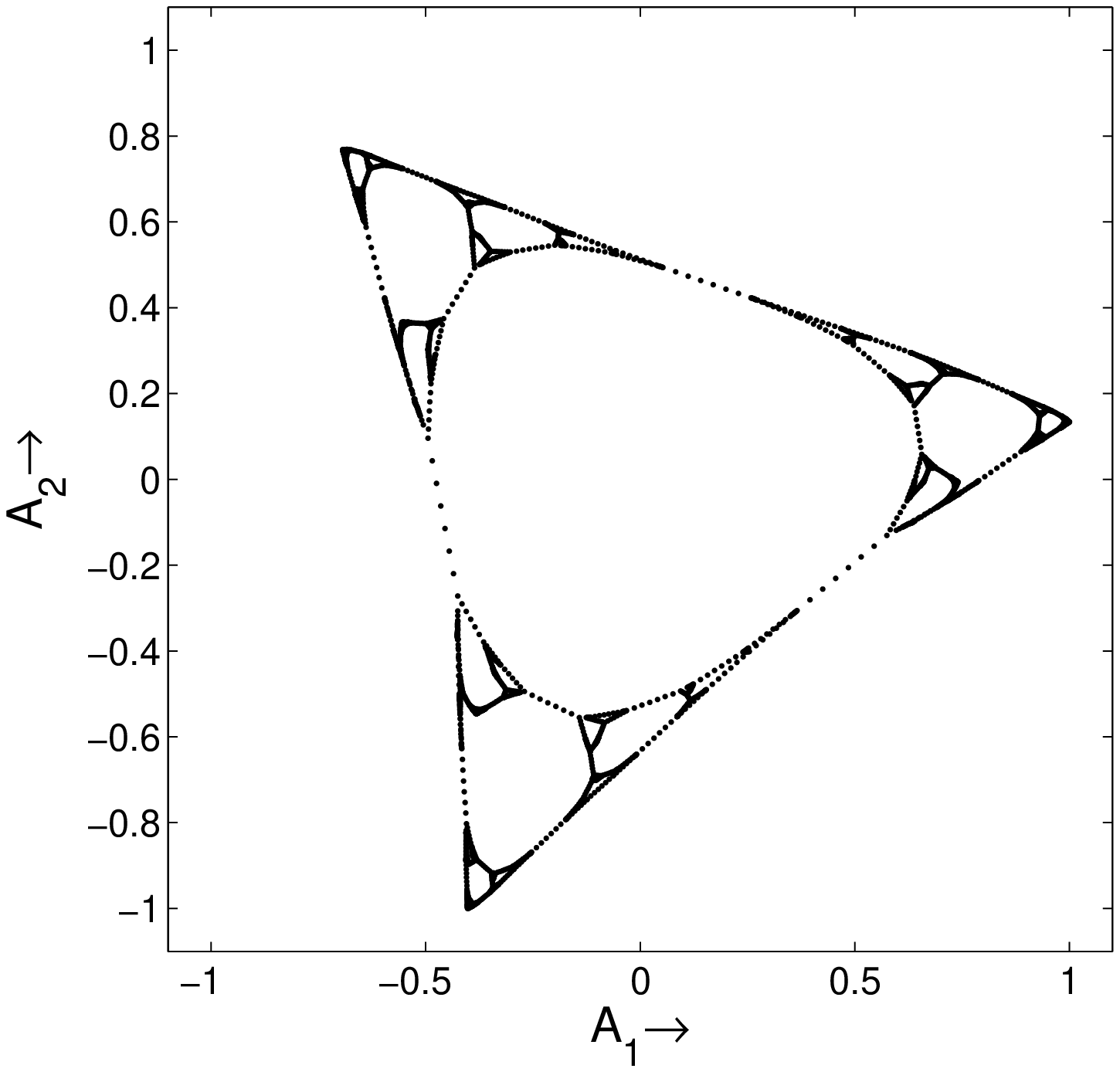}~
\includegraphics[height=.25\textheight,width=.3\textwidth]{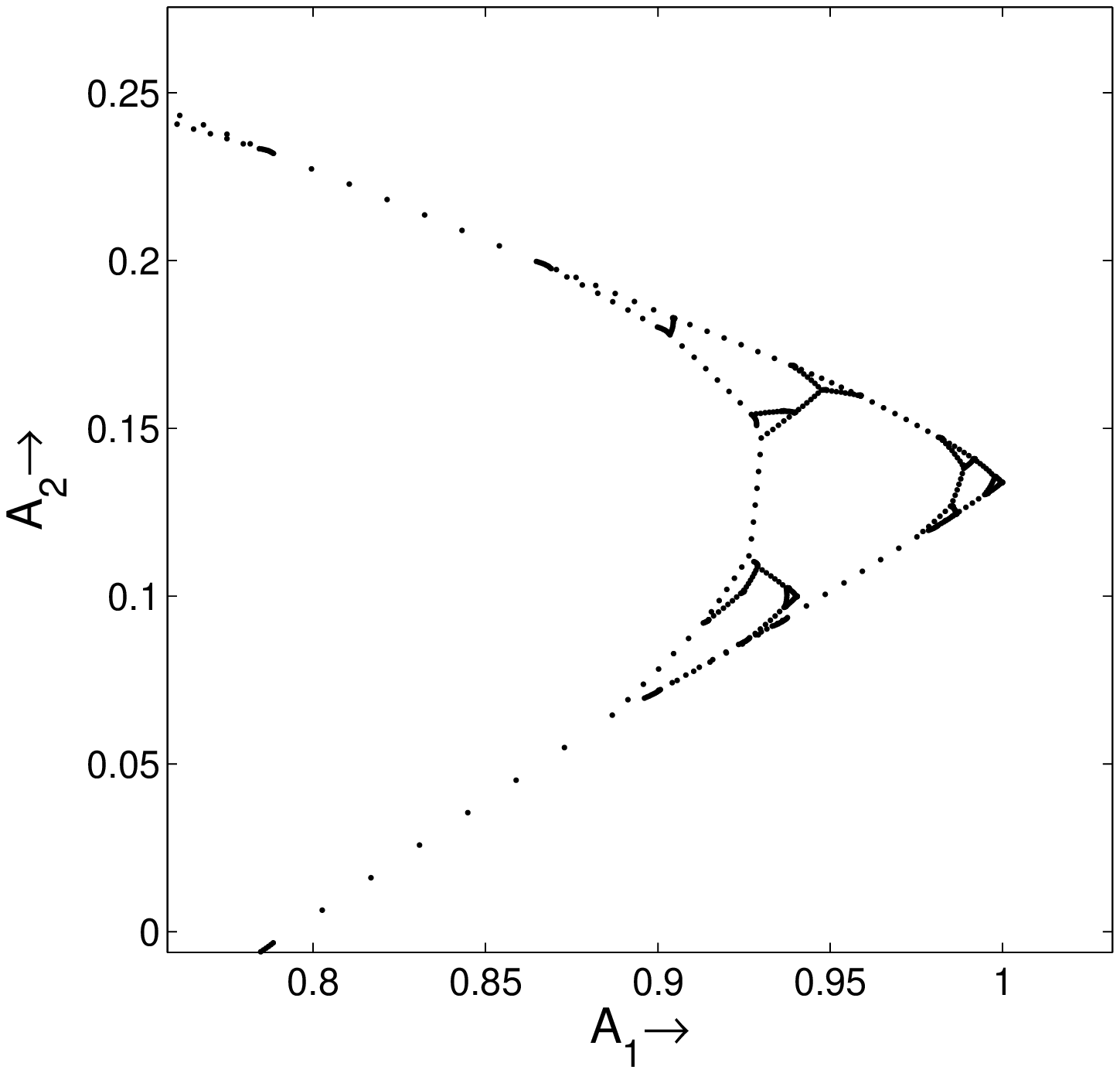}}
\caption{Observable representation for Brownian motion on a fractal built of triangles. This is \textbf{not} an image of the fractal coordinate space per se. Rather it is constructed from the eigenfunctions for the stochastic dynamics, as described in the text. On the right is a closer look. The magnification can be deduced from the values on the coordinate axes. 
\label{fig:fractal}}
\end{figure}

We drop the requirement that $\lambda_{m+1}$ be small. This was briefly considered in \cite{multiplephases}, where we illustrated that for Brownian motion on a ring, the observable representation (for $m=2$) gave a ring, while for motion on a planar region bounded by a square the observable representation gave a square. It turns out that this is only the tip of the iceberg: faithful imaging is a property of a large collection of dynamical systems, as suggested by \cite{grains} and demonstrated in \cite{coifman}. Fig.\ \ref{fig:fractal} shows the much more complicated observable representation for Brownian motion on a fractal structure built of triangles. This image is \textbf{not} a picture of the fractal per se, but is the set of points $(A_1(x),A_2(x))$  for $x\in X$, a discrete realization of the fractal, and $A_k$ eigenfunctions of the associated stochastic dynamics.

The reason this works is that \textit{dynamical proximity induces proximity in the observable representation}. Many indications of this were given in \cite{multiplephases}. The general idea however contains a subtlety: dynamical proximity in general is not symmetric (getting from $x$ to $y$ may be easier than getting from $y$ to $x$), while in the observable representation distances are necessarily symmetric. The inequality we will derive in Sec.\ \ref{sec:rationale} provides a relation to a symmetric dynamical construct. On the other hand when asymmetry exists it is often the marker of a particularly interesting class of states. We intend to discuss this in our study of the Sherrington-Kirkpatrick spin glass model \cite{skobservrep}.

\section{\label{sec:illustrationsone}Brownian motion on $V$: Imaging $V$}

\subsection{\label{sec:oneD}One dimension}

As indicated, for Brownian motion on a circle the $m=2$ observable representation gives a circle. This can easily be checked analytically. The transition matrix (or its generator $W$, related to $R$ by $W=(R-1)/\Delta t$) becomes, in the continuum limit, the Laplacian. In the one-dimensional case, the slowest eigenfunctions are sine and cosine (with degenerate eigenvalues) and a circle is obtained.

To see what happens for Brownian motion on a line one can also go to the continuum limit, where it is necessary to determine the boundary conditions. The fact that our discrete process conserves probability, meaning that there is reflection at the boundaries, implies the use of Neumann boundary conditions at the ends of the line. The eigenfunctions are essentially cosine and cosine of twice the argument. Plotted against each other, these give a line. The line is not straight, but the topology of the space is preserved. (See Fig.\ \ref{fig:BMLine} for an image corresponding to the discrete walk on a line of 100 points.)

\begin{figure}
\includegraphics[height=.4\textheight,width=.6\textwidth]{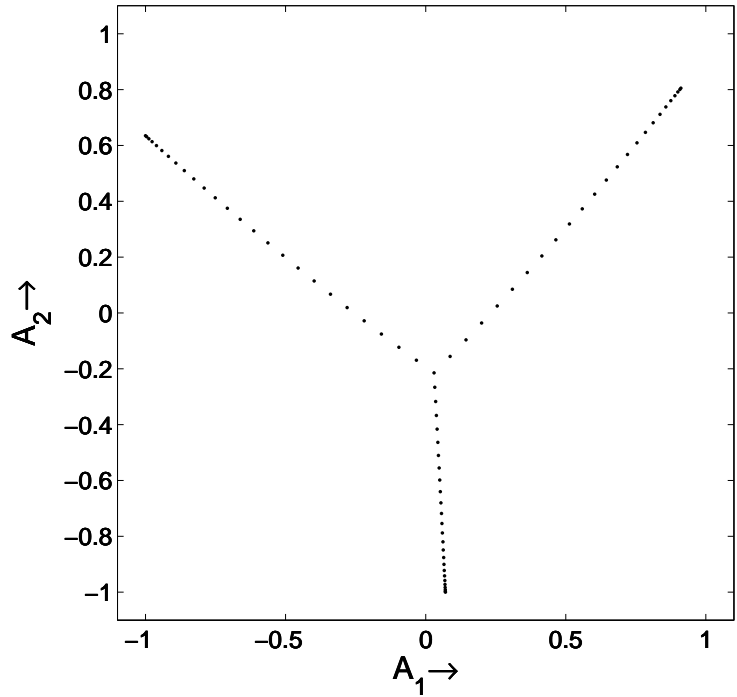}
\caption{Observable representation for Brownian motion on a figure-Y.\label{fig:Y}}
\end{figure}

\begin{figure}
\centerline{
\includegraphics[height=.3\textheight,width=.4\textwidth]{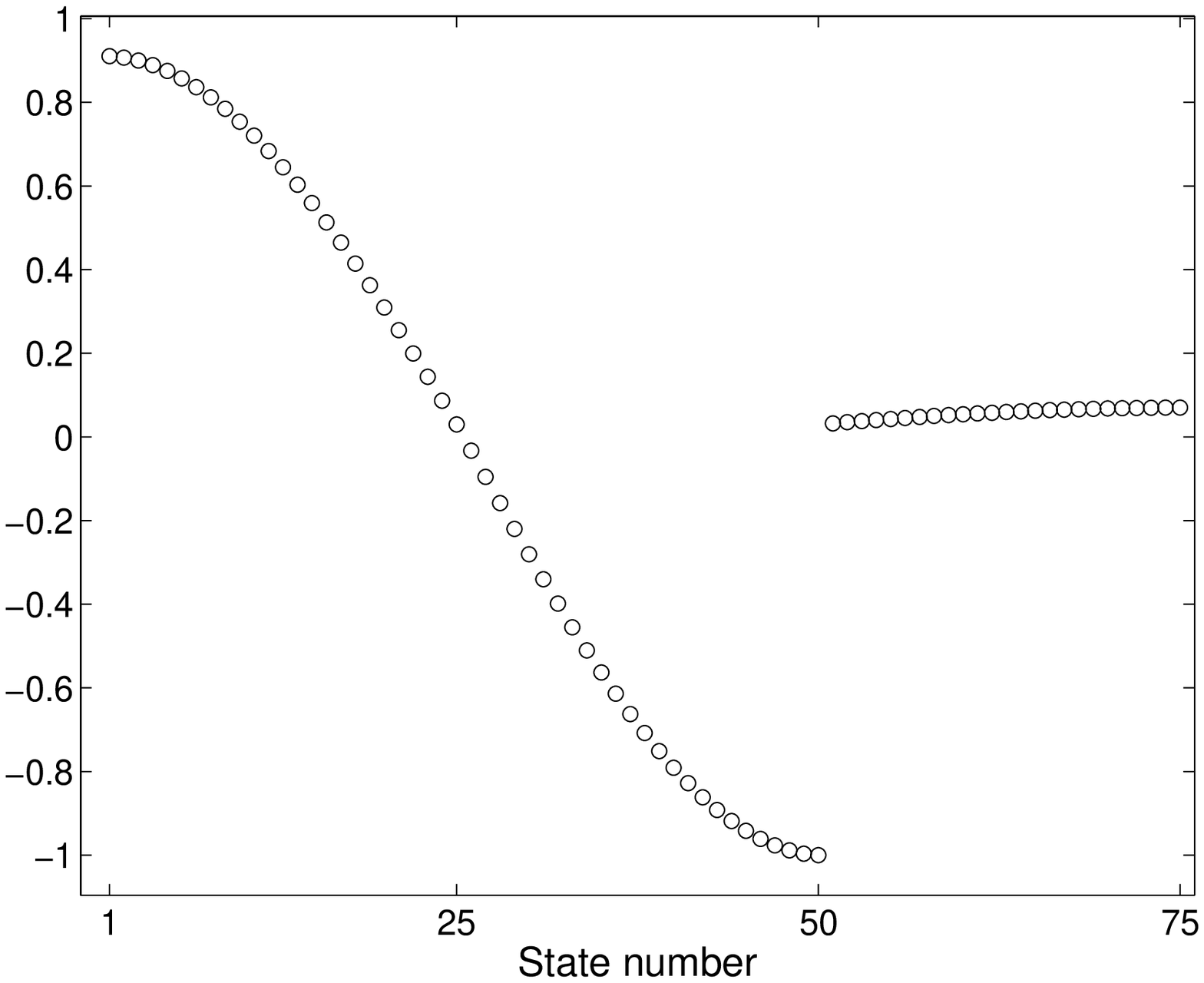}
\includegraphics[height=.3\textheight,width=.4\textwidth]{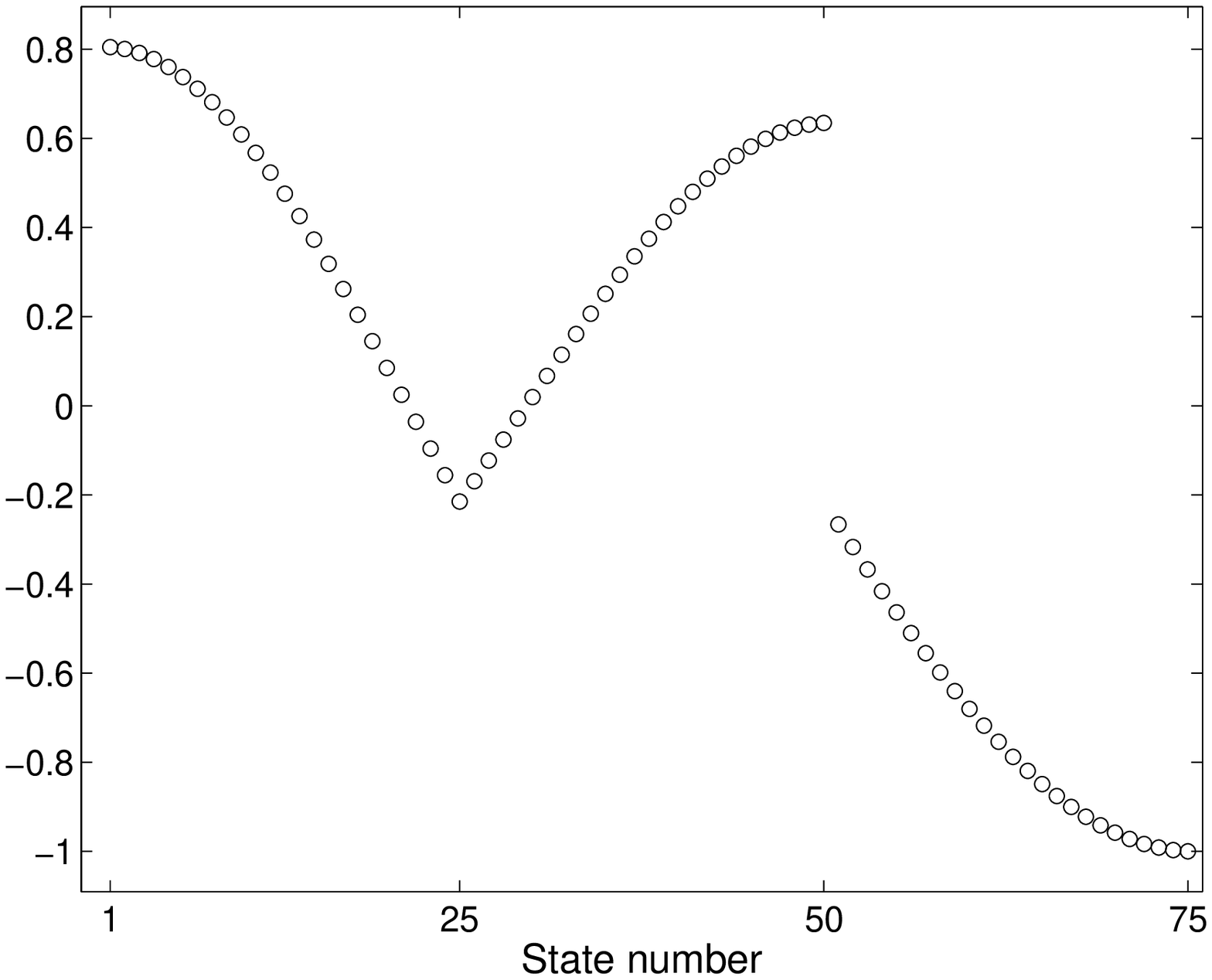}}
\caption{Eigenfunctions on which Fig.\ \ref{fig:Y} is based. On the left is the slowest eigenvector (eigenvalue of $W= -0.000985$). On the right is the next slowest (eigenvalue $-0.00141$). The fact that the trigonometric functions need not be the same on different branches leads to the crowding of points near the ends of the lines in Fig.\ \ref{fig:Y}.
\label{fig:Yeigvec}}
\end{figure}

For the next example we consider Brownian motion on a figure ``Y,'' three lines that meet at a point. In Fig.\ \ref{fig:Y} is the $A_1$-$A_2$ plot. Clearly the shape has been reproduced. To see how this has come about we show in Fig.\ \ref{fig:Yeigvec} a plot of the eigenfunctions. The points (states) are numbered so that 1, 50 and 75 are the endpoints, while state-51 is in contact with state 25. Note that besides the functions being continuous (the apparent jump at 50 is an artifact of the numbering scheme) there is a condition on the derivatives at the vertex (we now discuss the continuum limit). Specifically, for each $A_k$ the sum of the quantities $\partial A_k/\partial x$ must vanish at each vertex. This follows from integrating the Laplacian (in the general case, for which these considerations hold as well) across the vertex. Finally, observe that there is a discrete approximation to Neumann boundary conditions.

\begin{figure}
\includegraphics[height=.3\textheight,width=.45\textwidth]{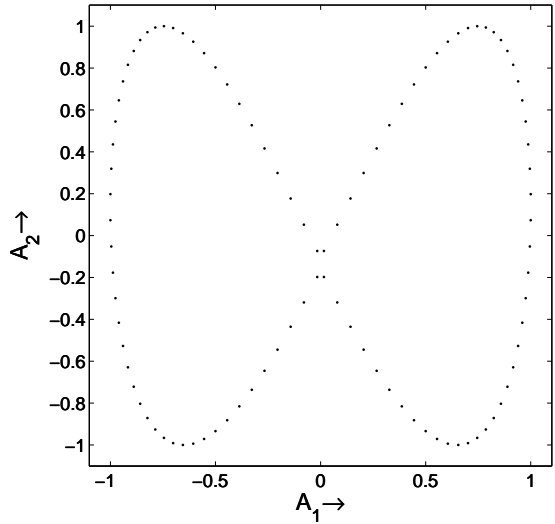}
\caption{Observable representation for Brownian motion on a figure eight .\label{fig:eight}}
\end{figure}

\begin{figure}
\centerline{
\includegraphics[height=.3\textheight,width=.4\textwidth]{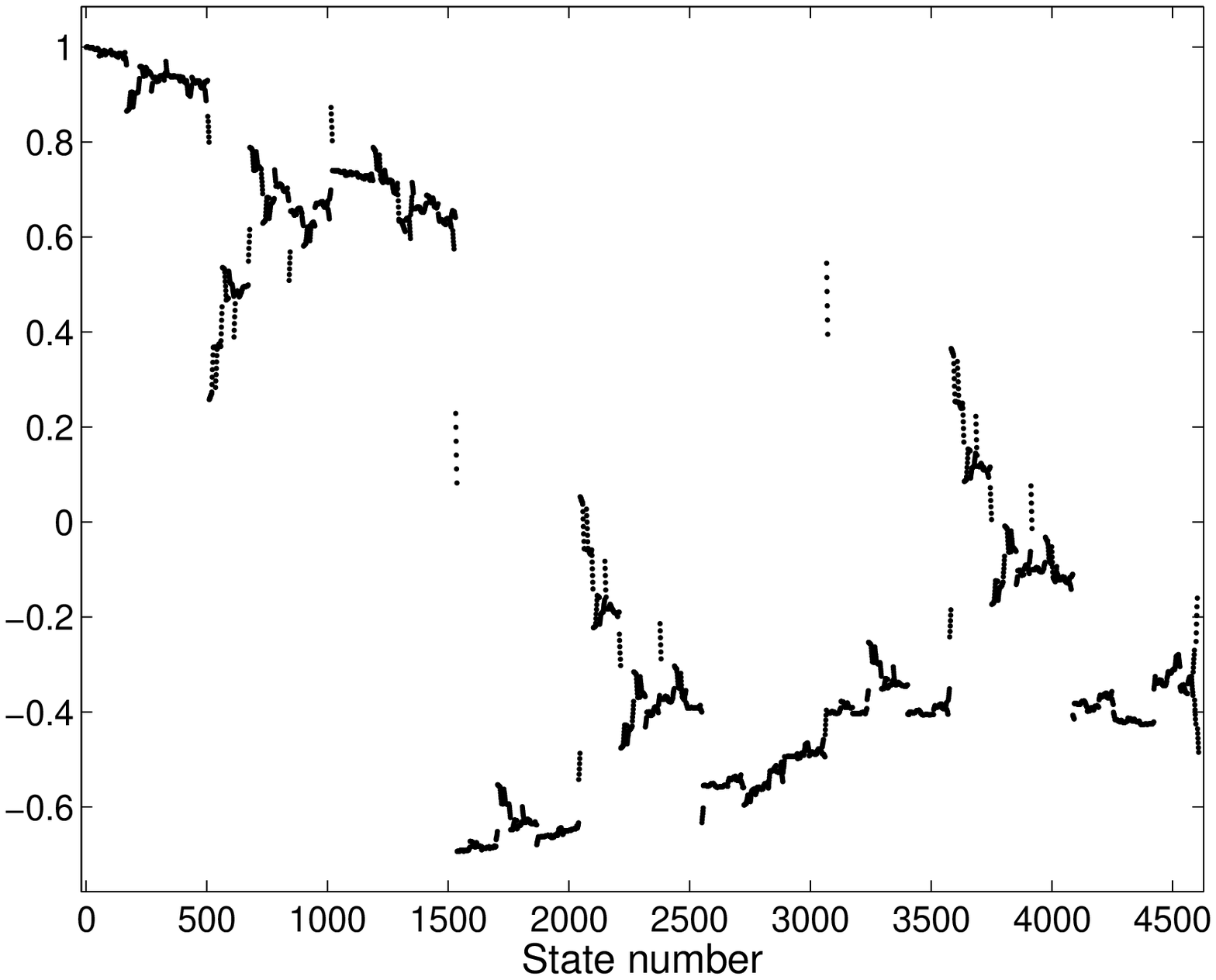}
\includegraphics[height=.3\textheight,width=.4\textwidth]{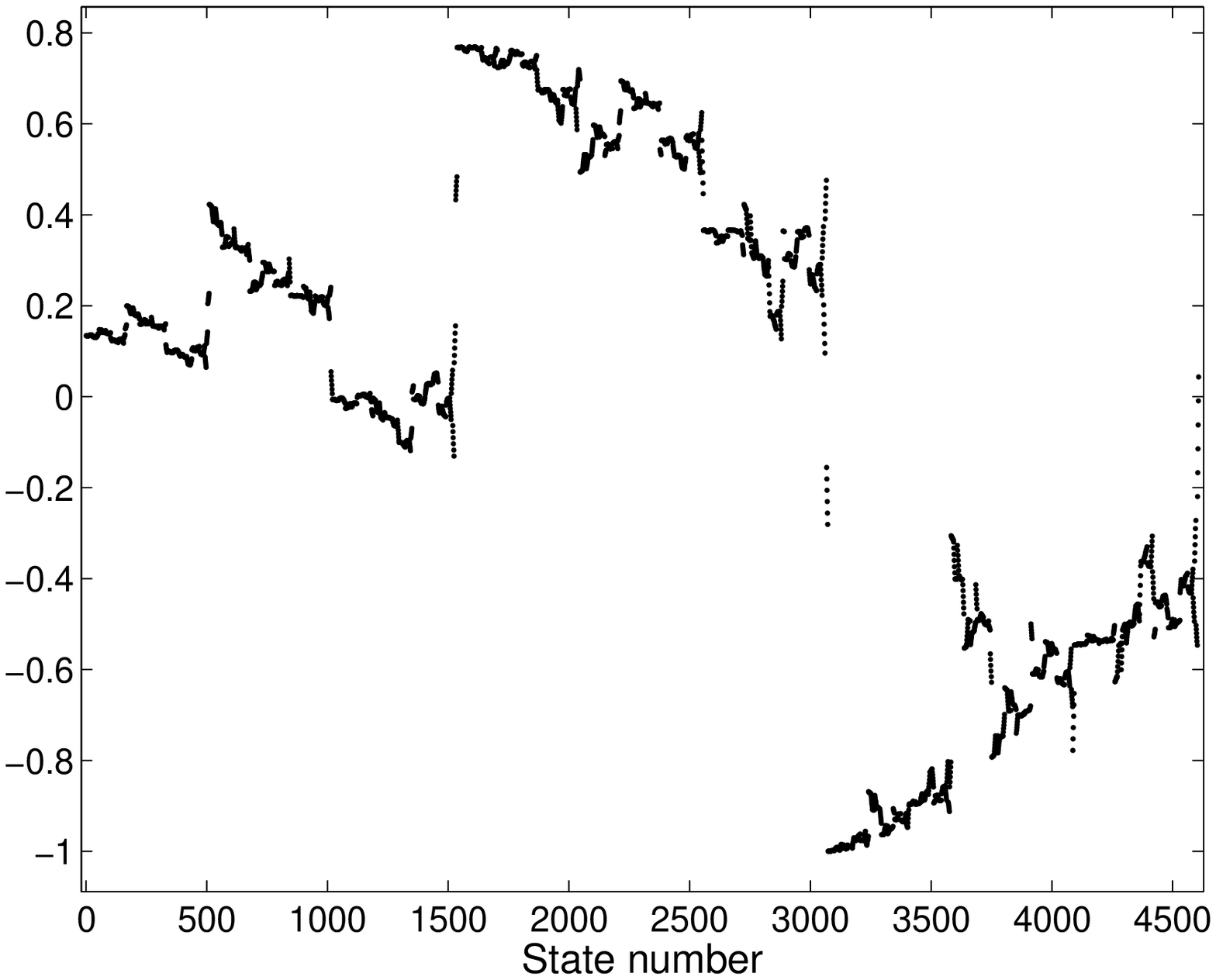}}
\caption{Eigenfunctions on which Fig.\ \ref{fig:fractal} is based.
\label{fig:fractaleigvec}}
\end{figure}

Brownian motion on a figure eight, Fig.\ \ref{fig:eight}, shows that again the topology of the coordinate space is preserved. This remarkable interplay of only two eigenvectors can be extended, as illustrated for the fractal in Fig.\ \ref{fig:fractal}. The eigenfunctions for the fractal, shown in Fig.\ \ref{fig:fractaleigvec}, produce a faithful representation of the coordinate space, although one would be hard-put to establish this by staring at the figure.

\begin{figure}
\centerline{\includegraphics[height=.3\textheight,width=.4\textwidth]{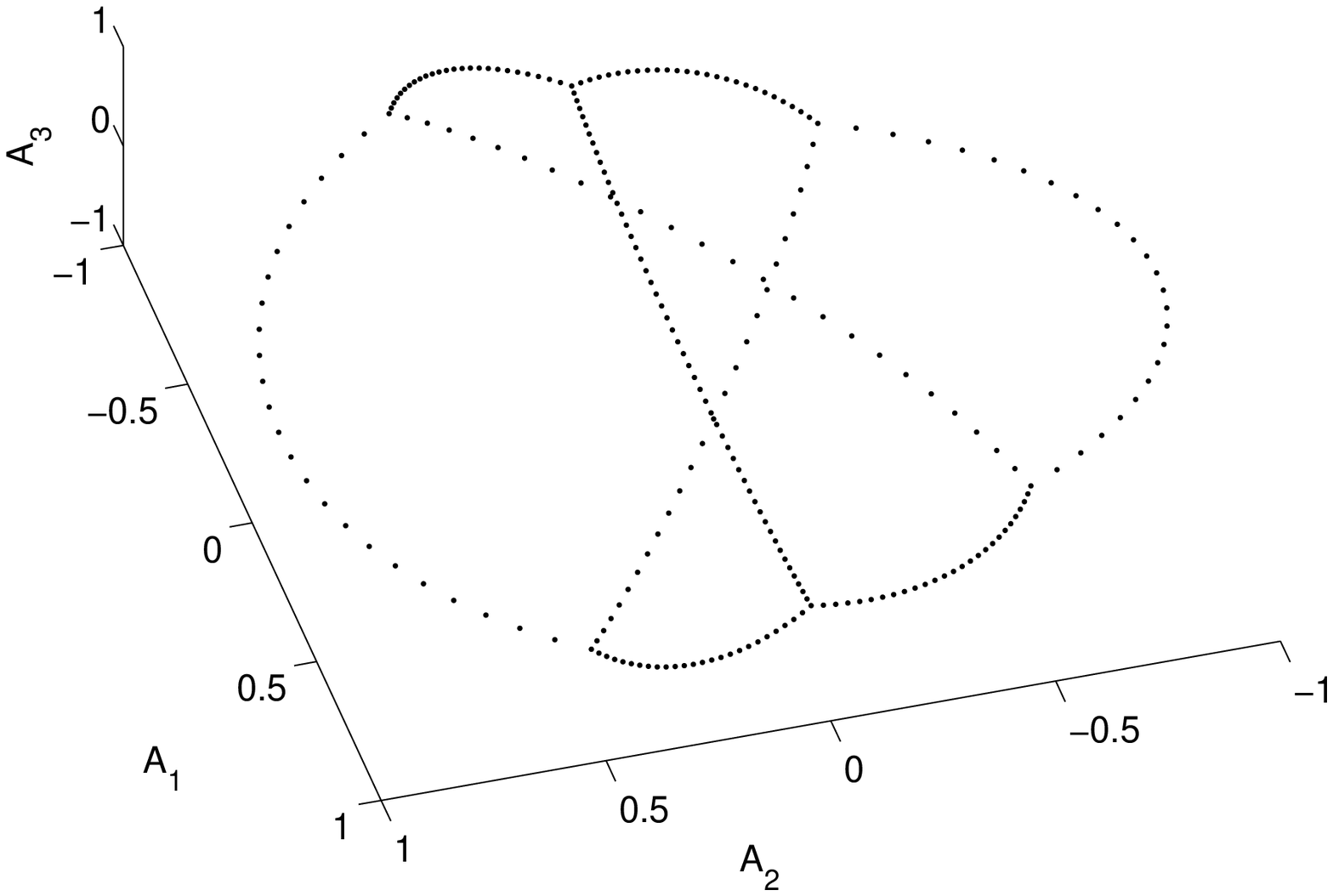}~
\includegraphics[height=.4\textheight,width=.6\textwidth]{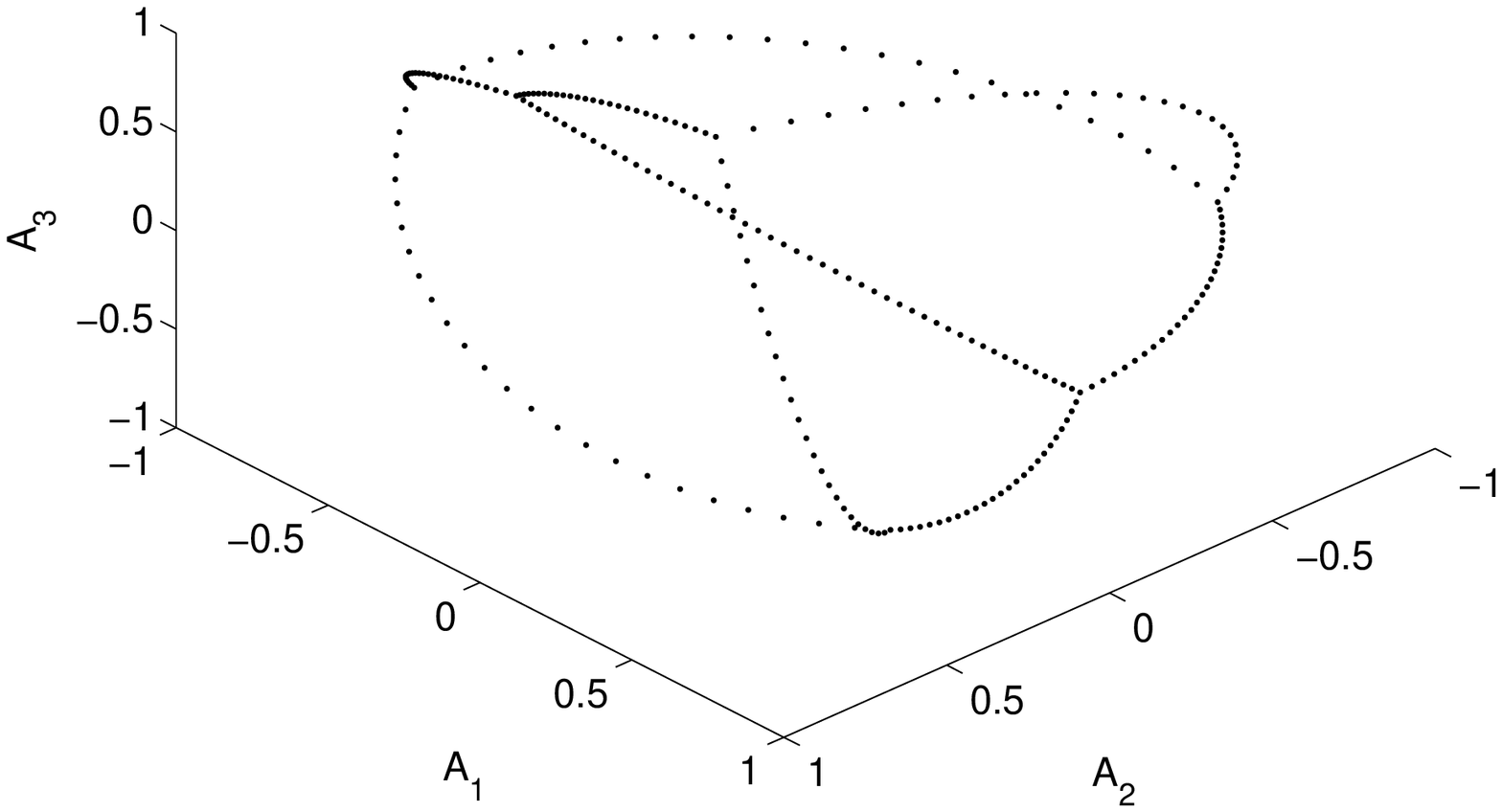}}
\caption{Two views of the observable representation for Brownian motion on a non-planar graph. The dimension ``$m$''  of the \obsrep\ must be set to 3 for fidelity. All vertices have 3 edges. The line crossings in both figures are artifacts of the two-dimensional projections, as can be ascertained by comparing the figures. \label{fig:nonplanar}}
\end{figure}

A further level of complexity is obtained for a non-planar graph (in this case K$_{33}$). Since the graph is indeed \textit{non}-planar, the $m=2$ \obsrep\ cannot accurately display the space. Fig.\ \ref{fig:nonplanar} shows two views of the 3-dimensional plot of $A_3$ versus $A_1$ and $A_2$.

Another non-planar graph, the Petersen graph, is known to have many unique properties \cite{holton, aigner}. For the observable representation it presents features not encountered in the examples given so far. As usual, we associate a matrix of transition probabilities with this 10-node graph by taking the adjacency matrix to be equal to (the non-diagonal part of) the generator, $W$, of the stochastic dynamics. $W$'s diagonal is fixed by $\sum_x W_{xy}=0$. (The generator is related to ``$R$'' by $R=1+W\Delta t$, with $\Delta t$ small enough for $R$'s matrix elements to be non-negative.) 

\begin{figure}
\centerline{\includegraphics[height=.35\textheight,width=.45\textwidth]{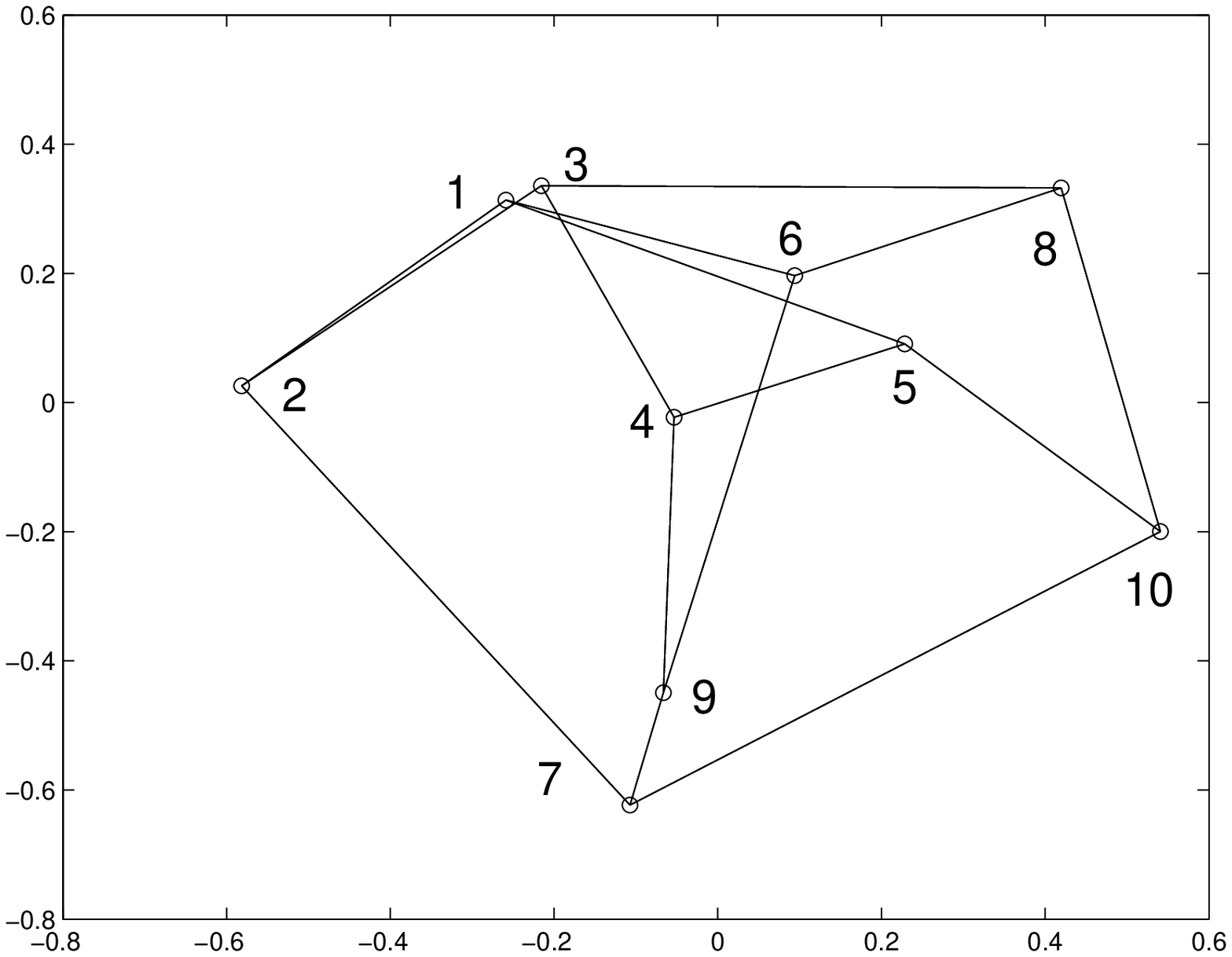}~
\includegraphics[height=.35\textheight,width=.45\textwidth]{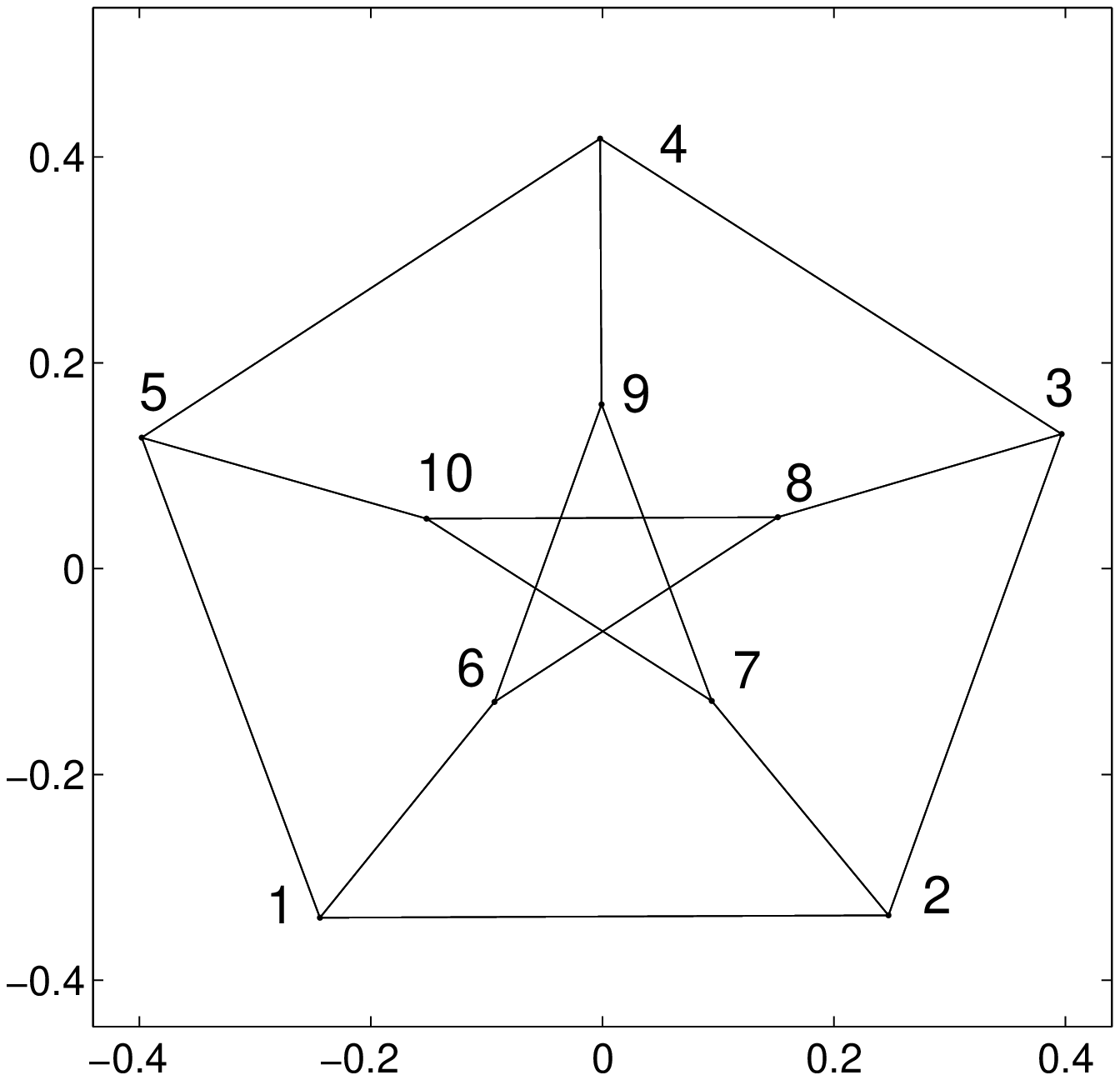}}
\caption{Observable representation for the Petersen graph. On the left is the result of a straightforward diagonalization using {\capsten matlab}. On the right the degenerate eigenvectors have been selected so as respect a five-fold symmetry of the graph. \label{fig:petersen1}}
\end{figure}

To illustrate the new feature present in the Petersen graph, in the left-hand image of Fig.\ \ref{fig:petersen1} we show the result of straightforward diagonalization and plotting. The lines drawn are not part of the observable representation, but only indicate which nodes are connected. The reason for this hodgepodge is that there are five degenerate eigenvalues, and which axes are chosen in this space depends only on vagaries of the diagonalization technique. In the same figure, on the right, we selected within this 5-dimensional space eigenfunctions that had particular properties with respect to the symmetries of the graph. The permutation (2,3,4,5,1)(7,8,9,10,6) commutes with $W$ and has eigenvalues that are the 5$^\mathrm{th}$ roots of unity. The real and imaginary parts of the complex simultaneous eigenvectors of $W$ and the permutation are then eigenvectors of $W$. These eigenvectors were used for the right-hand illustration in Fig.\ \ref{fig:petersen1}. This is one of the standard representations of the Petersen graph, and in particular makes manifest the 5-fold symmetry.

\subsection{\label{sec:manyD}More than one dimension}

\begin{figure}
\includegraphics[height=.4\textheight,width=.6\textwidth]{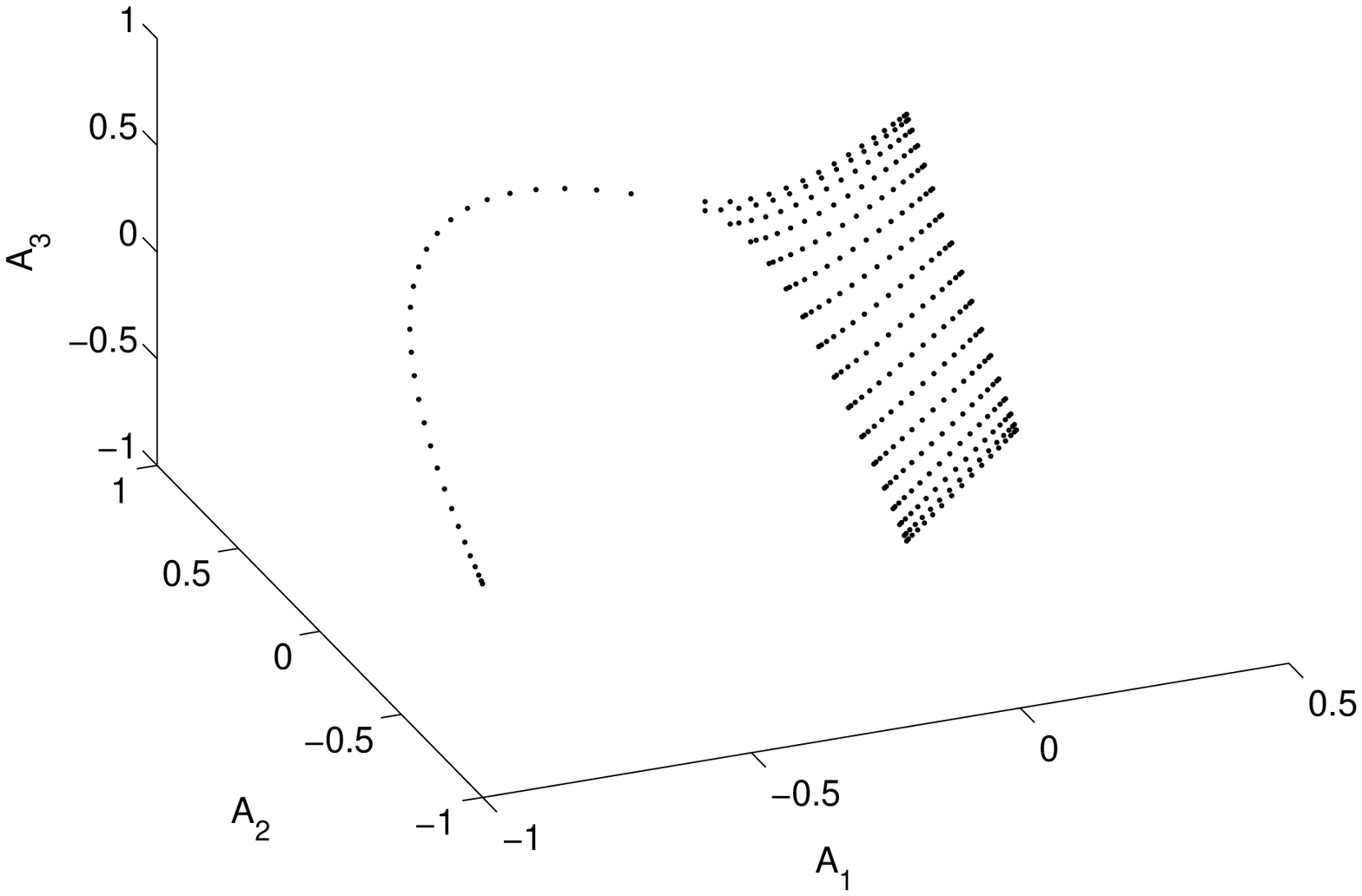}
\caption{Observable representation for Brownian motion on a square with a string attached at a corner.\label{fig:kite}}
\end{figure}

Our examples so far have dealt with one-dimensional coordinate spaces. But the method has no such restriction. As shown in \cite{multiplephases} the \obsrep\ for Brownian motion on a square is---simply---a square. In the next example we mix dimension. Suppose that there is a string attached to the square at one of its corners and the particle can wander off onto that string. Fig.\ \ref{fig:kite} shows the \obsrep\ for this stochastic process. We remark on a feature that can be seen in this figure as well as in Fig.\ \ref{fig:kite} and in Fig.\ 11 (on the right) in \cite{multiplephases}. There is a crowding of points near the edges. This is because one is plotting different trigonometric function against one another, and although no change in topology occurs, distances are not preserved.

\begin{figure}
\includegraphics[height=.4\textheight,width=.6\textwidth]{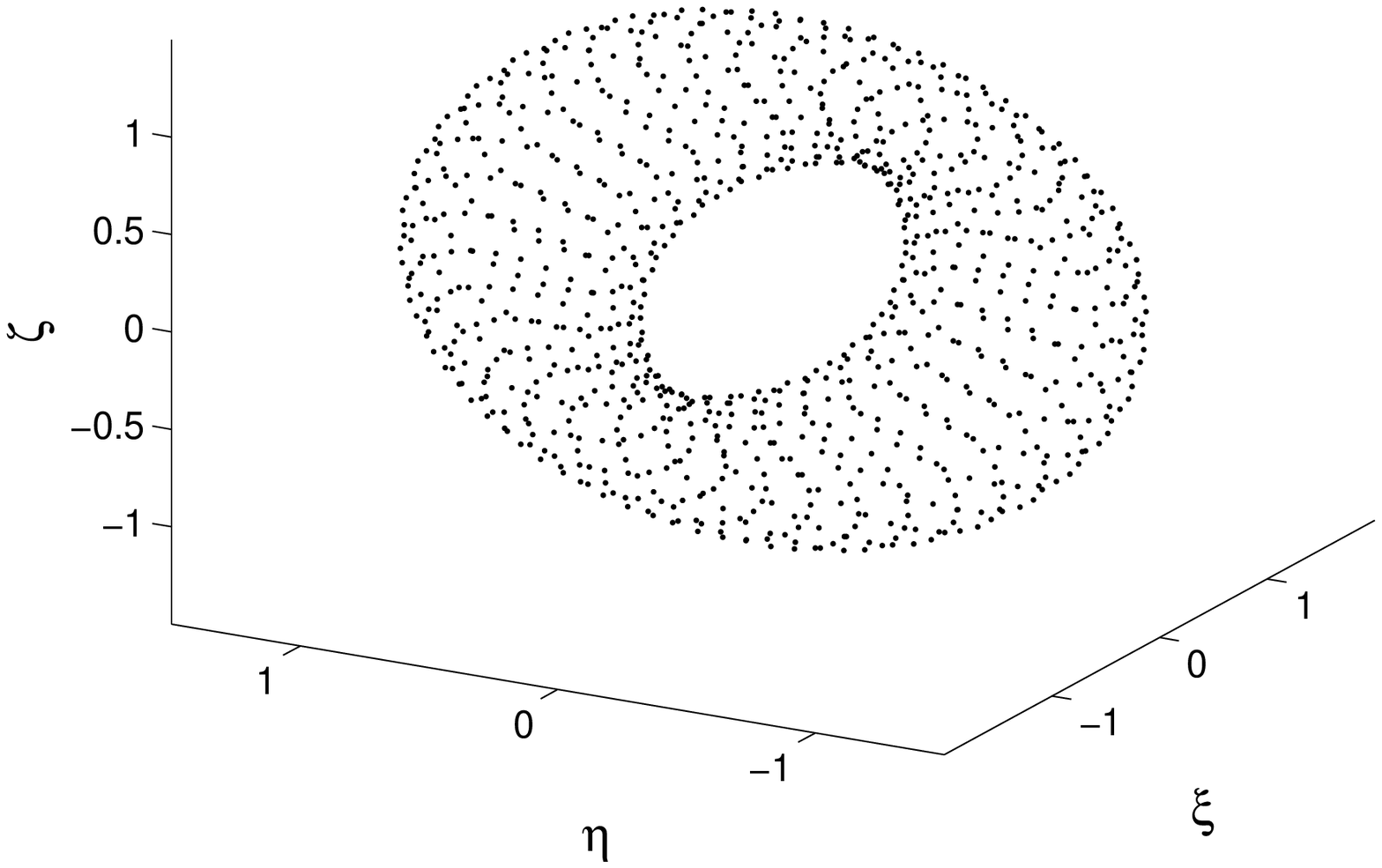}
\caption{Brownian motion on a torus. The axes are three random orthogonal vectors in the 4-dimensional observable representation. \label{fig:torus}}
\end{figure}

\begin{figure}
\includegraphics[height=.4\textheight,width=.6\textwidth]{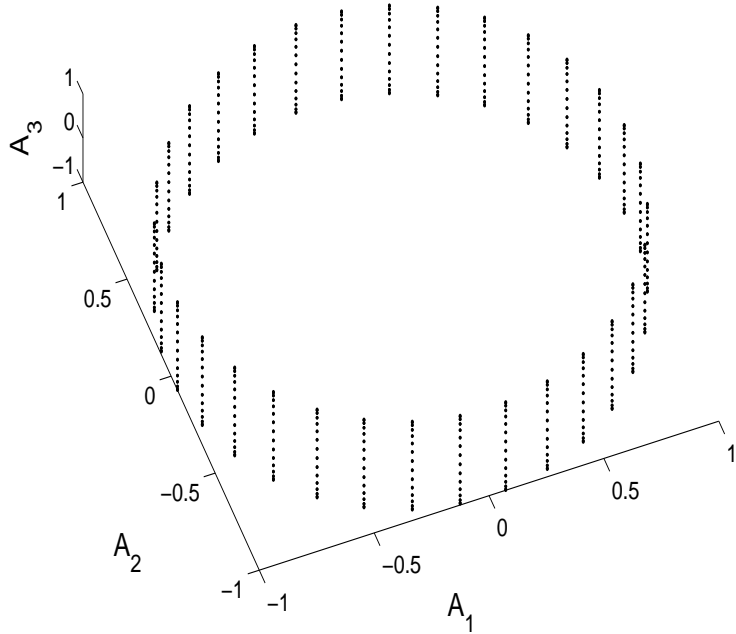}
\caption{Same as the previous figure, but the axes are $(A_1,A_2,A_3)$. \label{fig:badtorus}}
\end{figure}

A 2-dimensional space requiring embedding in three dimensions is the torus (i.e., we consider Brownian motion on the torus). This further illustrates a limitation of our method: when degeneracy appears, particular mixtures of the eigenvectors may cause the observable representation to collapse or project degrees of freedom of the coordinate space. (This problem was expressed a bit differently for the Petersen graph.) In Fig.\ \ref{fig:torus} we show a torus discretized at different levels for its two generating circles. For this situation, $\lambda_1=\lambda_2$ and $\lambda_3=\lambda_4$, but they do not equal each other. The figure shown is obtained by selecting a random 3-space within the four dimensions spanned by $(A_1,A_2,A_3,A_4)$ (and then projecting onto a plane for publication). However, if one takes as axes $A_1$ and $A_2$ and any direction in the plane $(A_3,A_4)$ one gets the image shown in Fig.\ \ref{fig:badtorus} (provided the discretization is different for the two radii: otherwise there is a four-fold degeneracy, effectively randomizing the axes).

Many other figures can also be constructed, for example the edge of one plane meeting a line through another. But these do not present problems and we omit them.

\section{\label{sec:illustrationstwo}Brownian motion on $2^V$: Imaging $V$\label{sec:kawasaki}}

Stochastic dynamics often describes the behavior of a collection of objects located in a space and having internal degrees of freedom whose interaction depends on the proximity of those objects within the space. An example is the Ising model with Glauber or Kawasaki dynamics. For this circumstance the path from dynamics to coordinate space is less direct. Nevertheless, since the dynamics \textit{does} contain spatial information, appropriate probes can lead to imaging of the coordinate space using the \obsrep.

\begin{figure}
\includegraphics[height=.4\textheight,width=.6\textwidth]
                      {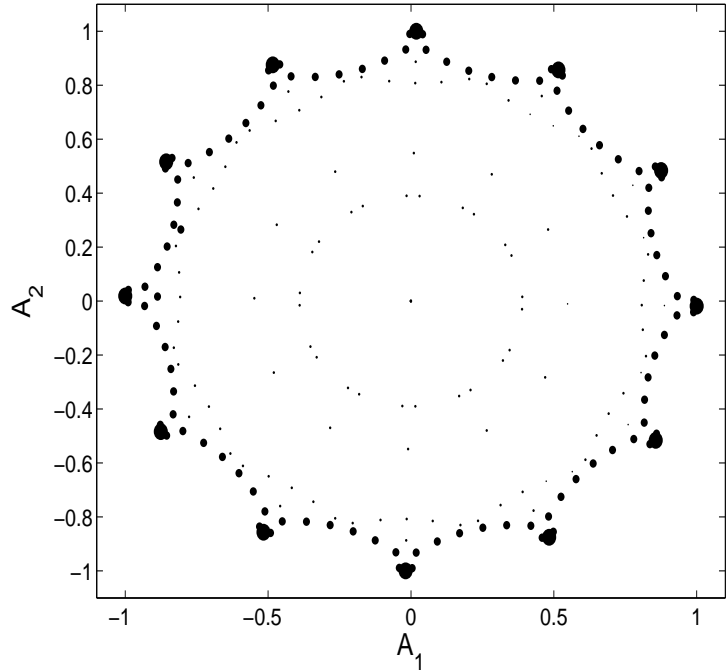}
\caption{Observable representation for the stochastic dynamics of nearest-neighbor Kawasaki dynamics on a ring of Ising (ferromagnetic) spins. Heavier symbols are used for states having greater $p_0$-weight. \label{fig:isingring}}
\end{figure}

In Fig.\ \ref{fig:isingring} we show the \obsrep\ for a ring of particles having ferromagnetic nearest neighbor interactions. We use Kawasaki dynamics at low temperature \cite{note:kawasaki}. The number of states is of course large, and because there is a conceptual distance between the space ``$V$'' and the dynamics on $2^V$ a bit of interpretation is in order. For this dynamics, the stationary state, $p_0$ is quite non-uniform and for those states having the largest $p_0$-weight we have used a larger symbol. We took 12 spins at low temperature with the condition that exactly 3 have the value +1. The 12 heavy points in the figure correspond the 12 possible clusters of 3 spins. They forms a circle because the stochastic process by which one can pass from one to the other of this states is ultimately a Brownian motion on a circle. That same symmetry exists for other possible spin configurations, 2 up, 1 down and all 3 separate. 

\begin{figure}
\includegraphics[height=.25\textheight,width=.4\textwidth]                     {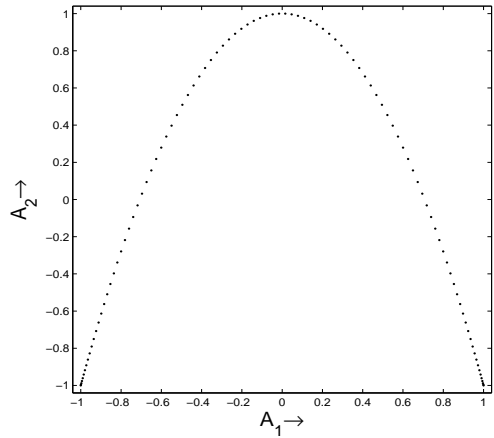}
\caption{Observable representation for Brownian motion on a line. The analytic form was given earlier; this image is presented for comparison with Fig.\ \ref{fig:isingline}. \label{fig:BMLine}}
\end{figure}

\begin{figure}
\includegraphics[height=.3\textheight,width=.6\textwidth]
                      {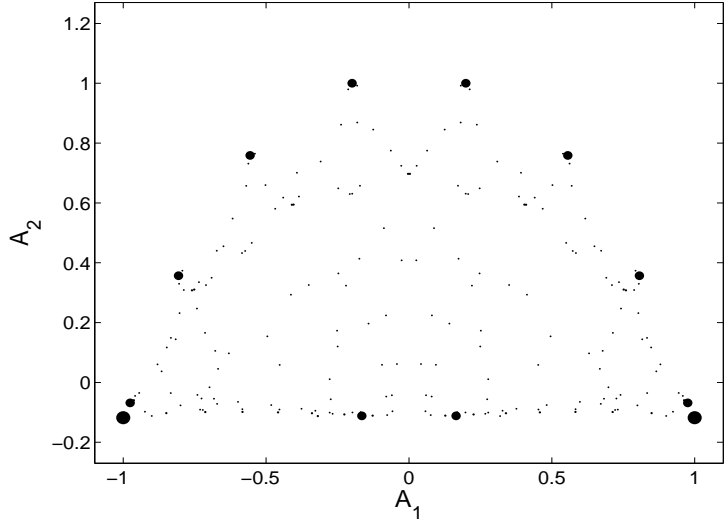}
\caption{Observable representation for the stochastic dynamics of nearest-neighbor Kawasaki dynamics on a line of Ising (ferromagnetic) spins. \label{fig:isingline}}
\end{figure}

A variation on this theme occurs for Kawasaki dynamics on a line. To orient the reader, we first show what the \obsrep\ gives for Brownian motion on a line. We earlier indicated how this is obtained analytically, but it is helpful to see the image. This is Fig.\ \ref{fig:BMLine}.
The \obsrep\ for Kawasaki dynamics on a line is given in Fig.\ \ref{fig:isingline}. The linear nature of the model emerges through the identification of points in the plot. The two points with maximal weight are on the lower left and lower right. These are clusters of three attached to the ends of the line. Because of the boundary condition they have only one broken bond and are favored over all other states. Clusters of 3 in the middle of the line require two broken bonds and are the next most likely states. These form the arc in the upper portion of the figure. This is because the transition from one to another is quite similar (if requiring more steps) to the simple walk on a line. Finally there are the two points near the bottom, between the line-end 3-clusters. They represent states with a cluster of two (up spins) on one end of the line and a cluster of one on the other. Because of the boundary condition, this costs only two broken bonds, giving these states the same weight as internal clusters of three. These are col points, in the sense that the quickest way to go from having one (maximal) cluster on the left, to one on the right, is via spin flips that take as intermediate values these two states. In line with the barycentric coordinate theorem of \cite{multiplephases}, their positions give the probabilities of reaching the minima. The shape of the arc associated with the random walk on the line is that same as what one gets from Brownian motion on a line, which is curved for reasons mentioned above. See Fig.\ \ref{fig:isingline}. There is a sharp intrinsic distinction between the intermediate points on the bottom and those of the arc (so this behaves quite unlike the circle shown earlier). As the number of spins changes, the two intermediate points on the bottom do not change. On the other hand, increasing (say) the number of spins will increase the number of heavy points on the arc.

\section{\label{sec:proliferate}Proliferating dimensions}

When a graph has many loose ends, the dimension required for a faithful representation may become large. We will look at two examples, the star and the Bethe lattice (complete binary tree). The difficulty is that the extremities of the graph have eigenfunctions that are entirely supported there, so that a few slow eigenvectors will not separate these points. The Bethe lattice is often described as being of infinite dimension, and this characterization is supported by the \obsrep\ as well.

A star consists of a central point surrounded by $n$ other points, all of which communicate \textit{only} with the central point. The generator of the stochastic dynamics is an $(n+1)$ by $(n+1)$ matrix whose only off-diagonal elements are ones in the first row and column. (The diagonal is adjusted so that $\sum_x W_{xy}=0$.) The eigenvectors to be used in the observable representation are $(n-1)$-fold degenerate. Calling them $\xi^{(\ell)}$, they have the property that their zero component is zero and the sum of components 1 through $n$ is zero. There are two other eigenvectors. One is constant on all points. The other is constant on the non-central points, but takes a different value on the central point. As a consequence the points of the \obsrep\ satisfy a single linear constraint but are otherwise independent. That constraint is that their sum is the zero vector, which is true because the components of the individual eigenvectors sum to zero.

We now present a specific construction showing that the \obsrep\ gives a simplex in $\erset^n$. Let $\nu\equiv n-1$. The unnormalized but orthogonal eigenvectors (reading horizontally, as in \Eqref{eq:visualization}) can be taken to be
\be
\ba{rrrrrrrrr}
0&1    & -1  &0    &0    &0    &0    &\dots&0  \\
0&1/2  &1/2  &-1   &0    &0    &0    &\dots&0  \\
0&1/3  &1/3  &1/3  &-1   &0    &0    &\dots&0  \\
0&1/4  &1/4  &1/4  &1/4  &-1   &0    &\dots&0  \\
\vdots                                         \\
0&1/\nu&1/\nu&1/\nu&1/\nu&\dots&\dots&\dots&-1 \\
\ea
\label{eq:starvectors}
\ee
There are $\nu=n-1$ rows in this array and $n+1$ columns. For the $(n+1)$ points of the \obsrep\ one reads down. The first column gives the (interior) point corresponding to the center of the star. We must show that the other $n$ points (i.e., columns 2 through $(n+1)$, considered as vectors in $\erset^n$) have non-zero volume. It is not difficult to show that the $\nu\times \nu$ matrix consisting of columns 2 through $n$ has determinant unity \cite{note:determinant1}.  This proves that these $\nu$ points have non-zero volume with respect to the origin of coordinates. To see that they have non-zero volume with respect to another point in $\erset^n$ one should subtract that point (a column vector) from every one of them, and evaluate the resulting determinant. Now from the determinant of columns 2 through $n$ one can subtract any sum of multiples of columns without changing its value. We therefore subtract the sum of \textit{all} the $\nu$ columns, which one notes is precisely the final column in the array. This proves our assertion \cite{note:secondproof}.

If one produces the \obsrep\ for any smaller dimension one still gets a simplex but it is not a full image of the graph: points will be coincident.

\begin{figure}
\includegraphics[height=.3\textheight,width=.6\textwidth]{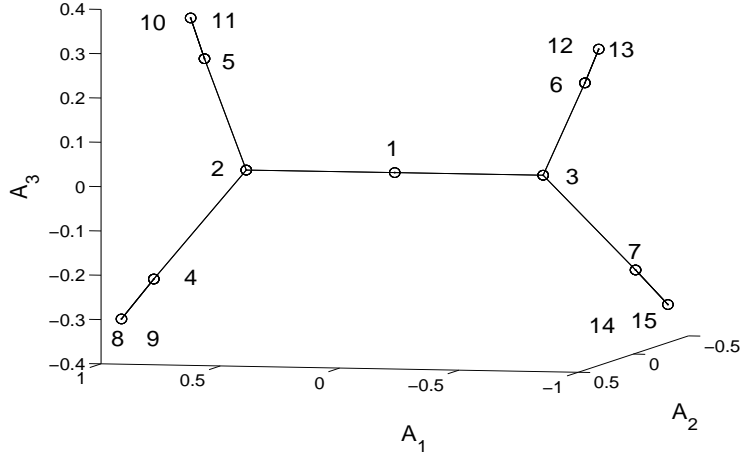}
\caption{Observable representation for a Bethe Lattice of 15 points. Note that points at the extremities of the graphs are not distinguished, reflecting the fact that the first three eigenvectors take the same values on these points. As for the star graph, increasing numbers of dimensions are required for a faithful representation. \label{fig:tree}}
\end{figure}

Similar results are obtained for the Bethe lattice. Here too there are eigenfunctions with support entirely on the fringes, so that omitting them yields a figure in which distinct points of the graph coincide in the \obsrep. In Fig.\ \ref{fig:tree} we show the result of a 3-dimensional plot of the 15 points of a tree with 3 levels of descendants. 

It is interesting that even if one fills in the lines of the graph, a faithful image is still not obtained. Edges of the graph were replaced by a sequence of points on which deterministic or Brownian motion took place. This made no substantial difference. On the contrary, this raises the question with respect to our Brownian motion figures: what is the minimal graph that would recover the same topological features. 

\section{\label{sec:rationale}Rationale }

In previous work we gave many relations of use in interpreting distance in the \obsrep\ \cite{multiplephases, creation}. The following (new) relation, \Eqref{eq:bound}, provides a general inequality. It is not particularly sharp, but applies throughout the set $\Acal$, not merely at its extremities. Our proof only applies when $R$ satisfies detailed balance.

Define $p_x(u,t)$ to be the probability distribution at time $t$ of a system whose time-0 position was $x\in X$. Thus $p_x(u,t)= R^t_{ux}$. A reasonable distance between two points $x$ and $y$ could be taken to be the variation distance between these two distributions. We look instead at something closely related:
\be
D(x,y;t)\equiv \sum_u\left|\frac{p_x(u,t)-p_y(u,t)}{\sqrt{p_0(u)}}\right| \,,
\ee
with $p_0$ the stationary state. By straightforward manipulation we rewrite $R$ in terms of $S$ and $\sigma$ (see Sec.\ \ref{sec:notation}) and use a spectral expansion for $S$. First note that
\be
\frac{p_x(u,t)}{\sqrt{p_0(u)}}
   =\frac{1}{\sqrt{p_0(u)}}\left(\sigma \Rs^t\sigma^{-1}\right)_{ux}
   =\sum_\alpha \lambda_\alpha^t \psi_\alpha(u)\frac{\psi_\alpha^\dagger(x)}{\sqrt{p_0(x)}}
   =\sum_\alpha \lambda_\alpha^t \psi_\alpha(u) A_\alpha(x)
\,,
\label{eq:pintermsofS}
\ee
where use has been made of \Eqref{eq:eigvecrelation} and the definitions preceding it. We next use \Eqref{eq:pintermsofS} combined with the fact that the L$^1$ norm exceeds the L$^2$ norm:
\be
D(x,y;t)\geq \sqrt{ \sum_u \left(\frac{p_x(u,t)-p_y(u,t)}{\sqrt{p_0(u)}}\right)^2}
=\sqrt{
\sum_u \left|\sum_\alpha\psi_\alpha(u) \lambda_\alpha^t \left[A_\alpha(x)-A_\alpha(y)\right]
\right|^2 
}
\,.\ee
This expression is of the form $\left[\sum_u\sum_\alpha\sum_{\alpha'}\psi_\alpha(u) \psi^*_{\alpha'}(u)c_\alpha c^*_{\alpha'}\right]^{1/2}$, which because of the orthonormality of the $\psi$'s is $\left[\sum_\alpha|c_\alpha|^2\right]^{1/2}$. It follows that
\be
D(x,y;t)\geq \sqrt{
\sum_\alpha |\lambda_\alpha|^t \left|A_\alpha(x)-A_\alpha(y)\right|^2
}
\,.\ee
The sum on the right can be truncated at the $m^\mathrm{th}$ eigenvalue, preserving the direction of the inequality. Since the magnitudes of the eigenvalues decrease monotonically (with index) we obtain, finally,
\be
D(x,y;t)/|\lambda_m|^{t}
\geq \sqrt{
\sum_{\alpha=1}^m \left|A_\alpha(x)-A_\alpha(y)\right|^2
}
\,.
\label{eq:bound}
\ee
The quantity on the right is the distance in the \obsrep. Thus, two points $x$ and $y$ which are adjacent dynamically---in the sense that the Markov chain soon nearly-forgets which point it came from---will be spatially adjacent in the \obsrep.

\section{Discussion\label{sec:discussion}}

Much of our previous work on the observable representation focused on metric properties. We used an assumed near degeneracy among the largest (slowest) eigenvalues to establish that points in separate phases formed tight clusters and were far from one another (in the \obsrep). In this article we drop the assumption of near-degeneracy. The information obtained thereby is therefore mainly topological. For this reason the bound obtained in Sec.\ \ref{sec:rationale} may be expected to be useful.

A second assumption that we drop is that \textit{after} the eigenvalues used in the \obsrep\ there is a significant decrease in the magnitude of the subsequent eigenvalues. From \Eqref{eq:bound} it is clear that one does not want to go to $m$ such that $|\lambda_{m}|$ is small, but \textit{not} having $|\lambda_{m+1}|$ be small presents no difficulties (whereas in \cite{multiplephases} it was important that there was a range of $t$ values for which $\lambda_m^t\approx1$, while $\lambda_{m+1}^t$ was small).

The message to be drawn from this work is that the topological features of an underlying space for stochastic dynamics are encoded in the eigenvectors. The \obsrep\ is a way to read that code. In some of our earlier work \cite{grains} that ``code'' was deciphered by creating clusters based on a dynamical metric (see also \cite{dynamicalmetric}). Here though we go directly to the eigenvectors. This same technique has been used by Coifman et al.\ \cite{coifman} and anticipated by Kodaira \cite{kodaira}. Our examples also deal with spaces that are neither manifolds nor graphs and also treat the recovery of geometry from a stochastic process in what is essentially a field defined on the underlying space. Finally, we remark that deducing spatial properties from the spectral properties of an operator defined on the space has a long history, with a well known example being Kac's ``Can one hear the shape of a drum?'' \cite{kac}. There however it is asymptotic properties of the eigenvalues that provide the information. Here it is the eigenfunctions of the slowest eigenvalues, functions that for physical reasons we term observables.

\begin{acknowledgments}
We thank Bertrand Duplantier, Thomas Gilbert, Peter Greiner, Annick Lesne, and Jean-Marc Luck for helpful discussions. This work was supported in part by NSF, NSA and ARO grants.
\end{acknowledgments}



\begin{thebibliography}{15}
\bibitem[{not({\natexlab{a}})}]{note:diag}
\bibinfo{note}{In using the operator ``$\diag$'' we follow {\capsten matlab}'s
  notation. Acting on a vector, $\diag$ creates a diagonal matrix with that
  vector on the diagonal, zeros elsewhere. Acting on a matrix, it produces a
  vector whose components are the diagonal of the matrix.}
\bibitem[{not({\natexlab{b}})}]{note:kawasaki}
\bibinfo{note}{In Kawasaki dynamics the random process allows (only)
  nearest-neighbor double spin flips for oppositely directed spins, hence
  preservation of the total magnetization.}
\bibitem[{not({\natexlab{c}})}]{note:determinant1}
\bibinfo{note}{One can show that the determinant is unity by induction
  (expanding about the right hand column of the matrix and doing successive
  expansions). Alternatively, recall that the rows in \Eqref{eq:starvectors}
  are orthogonal. Hence the determinant is the product of the Euclidean norms
  of the rows. Those are $\sqrt{1 + 1}, \sqrt{1 + 1/2},\sqrt{1 + 1/3},\dots$,
  except for the last row. Its value is $\sqrt{1/\nu}$ (for $\nu$ rows). The
  product is $\sqrt{2 \cdot \frac32\cdot\frac43\dots\frac\nu{\nu-1} \frac1
  \nu}$.}
\bibitem[{not({\natexlab{d}})}]{note:secondproof}
\bibinfo{note}{A second proof of the same assertion is as follows. First,
  none of columns $2,\dots,n+1$ in \Eqref{eq:starvectors} is within the
  convex hull of the others. Regarding columns 3 through $n+1$, each of
  them has a negative coefficient in a location different from that of 
  the others. As to column 2, it is not within the convex hull of the 
  others, because any convex combination of the remaining columns must 
  have a negative coefficient in the coordinate corresponding to the 
  first column used with positive weight. The same argument also shows 
  that any smaller subset of the columns are lower dimensional faces 
  of the larger simplex.}
\end{thebibliography}
\end{document}